\documentclass[12pt]{article}
\usepackage{epsfig}
 
\def\pmb#1{\setbox0=\hbox{#1}    % "poor man's bold"
 \kern-.025em\copy0\kern-\wd0    %  for bold \nabla
 \kern.05em\copy0\kern-\wd0      %
 \kern-.025em\raise.0433em\box0} %
\def\nbla{\pmb{$\nabla$}}        %
\font\amsfnt=msam10 % AMSTeX font. Use
\begin{document}
\date{}
\title{
Method of Envelope Functions and Intervalley $\Gamma$-$X_z$ Interaction
of States in (001) III-V Semiconductor Heterostructures}
\author{E.~E.~Takhtamirov$^{\dagger}$, V.~A.~Volkov$^{\ddagger}$
\\
Institute of Radioengineering and Electronics of RAS,
\\
Mokhovaya 11, 101999, Moscow, Russia
\\
$^{\dagger }$e-mail: takhtam@cplire.ru
\\
$^{\ddagger }$e-mail: VoVA@cplire.ru
}
\maketitle
%-----------------------------------------------------------------------

\begin{abstract}
The ${\bf k \cdot p}$ method is used to analyze the problem of intervalley $\Gamma$-$X_z$
interaction of conduction band states in (001) lattice-matched III-V semiconductor
heterostructures. A convenient basis for expansion of the wave function is systematically
selected and a multi-band system of equations is derived for the envelope functions which
is then reduced to a system of three equations for three valleys ($\Gamma_1$, $X_1$, and
$X_3$) by using a unitary transformation. Intervalley $\Gamma$-$X_z$ mixing is
described by short-range potentials localized at heterojunctions.
The expressions for the parameters determining $\Gamma$-$X_z$ mixing strength
explicitly contain the chemical composition profile of the structure, so mixing is
naturally stronger for abrupt heterojunctions than for structures with a continuously
varying chemical composition. It is shown that the direct $\Gamma_1$-$X_1$ interaction of
comparable strength to $\Gamma_1$-$X_3$ interaction exists. This must be taken
into account when interpreting tunnel and optical experiments since $X_1$
valley is substantially lower in energy than $X_3$ valley.
\end{abstract}
\vspace{0.3cm}

%-----------------------------------------------------------------------
\section{Introduction}

Heterostructures with quantum wells and barriers based on
GaAs/AlGaAs are popular objects for studying the physical processes accompanying
resonant tunneling and optical transitions. These structures are particularly
interesting because in GaAs the minimum of the conduction band is situated at the
center of the first Brillouin zone ($\Gamma_1$ symmetry) whereas the minimum of
the AlAs conduction band lies near $X$ point where the bands having
$X_1$ and $X_3$ symmetry are close together, $X_1$ band being the principal one
($X_3$ band is 350~meV higher energy). On each cubic axis, there is a pair of $X$
valleys, one from $X_1$ band and one from $X_3$ band.
Layers with predominantly Ga content act as barriers for $X$ electrons while regions
with Al predominating act as barriers for $\Gamma$ electrons. Under certain conditions,
resonant tunneling interaction can take place between energetically close states of
$\Gamma$ and $X$ valleys, and this is observed on the current-voltage
characteristics even of single-barrier structures (see, e.g., Refs.~\cite{teissier,dubrovskii}) and
also in optical experiments \cite{nakayama,ohtani}. The conditions for the existence of this
interaction generally involve the presence of structural defects, impurities, heterojunction
roughness, and/or interaction with short-wavelength phonons. However, in (001) heterostructures,
interaction of $\Gamma$ valley with $X_z$ valleys (i.e, $X$ valleys lying on the
quasi-momentum axis directed along the normal to the surface) is caused by
the potential of the structure itself which leads to non-conservation of the quasi-momentum
component $k_z$ perpendicular to the junction. As a result of size quantization for fairly
thin (less than 50 \AA ) AlGaAs layers, $X_z$-valley states are situated below all $X$
valleys so that an analysis of $\Gamma$-$X_z$ interaction at the heterojunction is
important for an accurate description of $\Gamma$-$X$ transitions \cite{raichev}.

This type of intervalley mixing has been studied theoretically both phenomenologically
\cite{liu} and using a tight-binding model \cite{lu,ando}, and also using a
pseudo-potential method \cite{franc}. Nevertheless, the level of understanding of the
processes leading to $\Gamma$-$X_z$ mixing of electronic states is far from satisfactory.
For instance, the results of Ref.~\cite{ando,franc} indicate that the direct $\Gamma_1$-$X_1$
interaction is extremely weak and $\Gamma$-$X_z$ mixing is merely attributable to
$\Gamma_1$-$X_3$ interaction, whereas, according to Ref.~\cite{lu}, $\Gamma_1$-$X_1$
interaction is the determining factor (however, the parameters of the tight-binding
model used in Ref.~\cite{lu} are such that $X_3$ band is very high in terms of energy
and is in fact eliminated from the analysis). Studies \cite{foreman,klipstein},
in which the form of $\Gamma$-$X_z$ interaction potential was determined by direct
calculations of the matrix elements of the model heterointerface potential using the
complete wave functions of the states also give a contradictory answer to the question
of the strength of $\Gamma_1$-$X_1$ interaction. According to the results of
Ref.~\cite{foreman}, this interaction is weak whereas the results of Ref.~\cite{klipstein}
suggest the opposite: the strength of $\Gamma_1$-$X_1$ mixing can be comparable to those of
$\Gamma_1$-$X_3$ mixing. The solution of this problem is important for interpreting tunneling
and optical experiments since $\Gamma_1$-$X_1$ transitions are usually observed experimentally.

In the present study $\Gamma$-$X_z$ interaction (mixing) is analyzed using the method
of envelope functions. This is a fairly explicit technique which does not have the
disadvantage of the tight-binding model in which the heterointerface is oversimplified,
while calculations using the empirical pseudo-potential method are much more
cumbersome than those using the envelope function method. Note that so far attempts to
derive a system of equations for the envelope functions jointly describing
$\Gamma$ and $X$ states have not produced satisfactory results. This is mainly because
an atomically abrupt change in the crystal potential at the heterojunctions produces
$\Gamma$-$X_z$ mixing \cite{schulman}, and correct allowance for such abrupt changes in
the potential is outside the scope of the usual method of Luttinger-Kohn envelope
functions. In Refs.~\cite{foreman,klipstein}, in which the problem was analyzed using the
envelope function method, $\Gamma$-$X$ interaction was analyzed using perturbation theory, but
as a basis there was selected a generally over-full set corresponding to the set of
Kohn-Luttinger functions for $\Gamma$ and $X$ states. An over-full (and nonorthogonal)
basis can, in principle, give an erroneous result. In addition, this approach cannot be applied
directly to describe the states of the continuous spectrum which is important for
$\Gamma$-$X_z$ tunneling problem. Nevertheless, the results of the present study agree
qualitatively with the conclusions of Ref.~\cite{klipstein}.

The problem of adequate description of intervalley mixing of states in
heterostructures is similar to the problem of intervalley splitting of impurity states in
multi-valley semiconductors (for example, in Si and Ge). It is known, see Ref.~\cite{bassani},
that allowance for the short range part of the impurity potential (``the correction to
the central cell'') not only yields a ``chemical shift'' of the impurity-state energy but also
lifts the valley degeneracy. In Ref.~\cite{we1}, we proposed a fairly simple method of
analyzing heterojunctions with an atomically abrupt change in chemical composition.
This method involves isolating the ``smooth'' component of the heterostructure potential
and the ``sharp'' component which is only nonzero near the heterojunction. The smooth
component is ``processed'' by a standard technique (the Kohn-Luttinger method) while the
sharp component is considered as a correction to the central cell. In Ref.~\cite{we1}
we only considered states near $\Gamma$ point in the Brillouin zone and in the present
study we develop the method further to describe the interaction of states near
different points in ${\bf k}$ space. These results were first presented at the III
All-Russia Conference on the Physics of Semiconductors, see~\cite{wegx}, and also \cite{disser}.
%-----------------------------------------------------------------------

\section{Formalism of the Envelope Function Method}

\subsection{Formulation of the Problem}

We shall consider the electron states in (001) III-V heterostructures formed from related
lattice-matched (with the lattice constant $a$) semiconductors having zinc blende symmetry.
By related structural materials we understand fairly small band offsets so that in the energy
range of interest the conduction band states near $\Gamma $ point can be
described using a single-band variant and the states near the point $X$ can be described
using two-band (for $X _1$ and $X _3$ bands) variants of the envelope function
method. We shall also assume that the energy gap between the states of interest to us in
the $\Gamma$ and $X$ valleys is smaller than or of the order of the band offset.
This means that we can consider the direct interaction of $\Gamma _1$, $X _1$, and
$X _3$ states ``exactly'' and interaction via all other bands can be taken into account
using perturbation theory. Since the desired Hamiltonian of the equation for the envelope
functions should depend explicitly on the number of monoatomic layers of each
material forming the structure \cite{lu,ivchenko}, we shall consider a structure with
two symmetric heterojunctions as the simplest case to obtain this dependence. For simplicity,
we shall neglect spin-orbit interaction and also external smooth potentials.
The single-electron Schr\"odinger equation then has the following form:
%====================================================================
\begin{equation}
\left( \frac{{\bf p}^2}{2m_0}+ U \right)
\Psi \left( {\bf r}\right) =\epsilon
\Psi \left( {\bf r}\right) . \label{schroedinger}
\end{equation}
%=====================================================================
Here $m_0$ is the free electron mass, ${\bf p}$ is the momentum operator, and
$U\equiv U\left( {\bf r}\right)$ is the crystal potential of the heterostructure.
We shall first use the following model for this potential (a more realistic situation will be
discussed in Section 3):
%======================================================================
\begin{equation}
U=U_1+P\left( z\right)\left[ U_2-U_1\right] \equiv
U_1+P\left( z\right) \delta U,
\label{lattice}
\end{equation}
%=======================================================================
and $U_1\equiv U_1\left( {\bf r}\right)$ and $U_2\equiv U_2\left( {\bf r}\right)$
are periodic (continued to all space) potentials of the two heterostructure materials,
the $z$-axis is perpendicular to the heterojunction plane; the form-factor $P(z)$ of the
heterostructure having the heterointerfaces at $z=0$ and $z=L$ is defined so that
%=======================================================================
\begin{equation}
P\left( z\right)=\cases{0,& $z<-d$; \cr 1,& $d<z<L-d$; \cr 0,& $z>L+d$.}
\label{form}
\end{equation}
%=======================================================================
The behavior of the function $P(z)$ in the transition regions (of width $2d$ each)
near the heterointerfaces may be fairly arbitrary. Here, the symmetry of the structure
implies that $P(z)=P(L-z)$, $L>d$. We shall assume that the layer of width $L$ contains an
integer number of monolayers: $L={\cal N}a/2$, where $\cal N$ is a natural number.
%-----------------------------------------------------------------------

\subsection{Choice of Complete Orthonormalized Set of Basis Functions}

As in Ref.~\cite{we1}, the potential $P(z)\delta U$ is analyzed in terms
of perturbation theory and the complete and orthonormalized set of functions used to
expand the complete wave function $\Psi ({\bf r})$ is constructed of Bloch functions of
the base semiconductor (having the crystal potential $U_1$). In our case, the most natural
basis is a mixed basis of Kohn-Luttinger functions for the points $\Gamma$ and $X _z$. In
order to make this set complete and orthonormalized, it should be constructed as follows.
We first expand $\Psi ({\bf r})$ in terms of the Bloch functions
$u_{n{\bf k}}({\bf r})e^{i{\bf kr}}$ of the base crystal which correspond to the energy
eigenvalues $\epsilon_{n{\bf k}}$, where $n$ and ${\bf k}$ are the band index and the
quasi-momentum vector, respectively:
%======================================================================
\begin{equation}
\Psi \left( {\bf r}\right) =\sum_{n^{\prime }}\int {\cal A} _{n^{\prime }}
\left( {\bf k}^{\prime }\right) e^ { i{\bf k}^{\prime } {\bf r}}
u_{n^{\prime }{\bf k}^{\prime}} \, {\rm d}^3k^{\prime }.  \label{expans1}
\end{equation}
%======================================================================
Summation in (\ref{expans1}) is performed over all bands and integration is performed
over the region $\Lambda_0$ nonequivalent ${\bf k}$. In order not to consider two equivalent
$X_z$ points (having the coordinates $(0,0,2\pi/a)$ and $(0,0,-2\pi/a)$ in ${\bf k}$ space),
we shall not operate in the first Brillouin zone constructed as a Wigner-Seitz cell but we
shall define a region $\Lambda_0$ such that points $\Gamma$ and, for example,
${\bf q}=(0,0,2\pi/a)$ are contained in this region with their vicinities.
Following Ref.~\cite{bassani} (Sec. 7-3), we divide $\Lambda_0$ into two subregions
$\Lambda_{\Gamma}$ and $\tilde \Lambda _X$ containing the points $\Gamma$ and $X_z$ with their
vicinities, where $\Lambda_{\Gamma} \cup \tilde \Lambda_X =\Lambda_0$ and
$\Lambda_{\Gamma} \cap \tilde \Lambda_X =0$ (see comments on this method of division in
Section 2.4). Now, following Ref.~\cite{lut-kohn} we use series expansions of the
periodic function $u_{n{\bf k}}$:
%========================================================================
\[
u_{n{\bf k}}=\sum_{m^{\prime }}b_{m^{\prime }n}\left( {\bf k}\right)
u_{m^{\prime }0}, \quad
u_{n{\bf k}}=\sum_{m^{\prime }}c_{m^{\prime }n}\left( {\bf k}\right)
u_{m^{\prime }{\bf q}}.
\]
%========================================================================
Then (\ref{expans1}) can be rewritten in the following form:
%======================================================================
\[
\Psi \left( {\bf r}\right) =\sum_{n^{\prime },m^{\prime }}
\int_{{\bf k}^{\prime}\in \Lambda_{\Gamma}} {\cal A}_{n^{\prime }}
\left( {\bf k}^{\prime }\right) b_{m^{\prime }n^{\prime}}
\left( {\bf k}^{\prime}\right) e^ { i{\bf k}^{\prime }
{\bf r}} u_{m^{\prime }0} \, {\rm d}^3k^{\prime }+
\]
\begin{equation}
+\sum_{n^{\prime },m^{\prime }}\int_{{\bf k}^{\prime} + {\bf q}\in \tilde
\Lambda_X} {\cal A}_{n^{\prime }}\left( {\bf k}^{\prime }+{\bf q} \right)
c_{m^{\prime }n^{\prime}}\left( {\bf k}^{\prime} + {\bf q}\right)
e^{i{\bf k}^{\prime } {\bf r}} e^{ i{\bf q}{\bf r}}
u_{m^{\prime }{\bf q}} \, {\rm d}^3k^{\prime }. \label{expans2}
\end{equation}
%======================================================================
We then define the functions 
%======================================================================
\begin{equation}
{\cal F}^{\left(\Gamma\right)}_{m^{\prime }}\left( {\bf k}^{\prime}\right)
=\sum_{n^{\prime }} {\cal A}_{n^{\prime }}\left( {\bf k}^{\prime }\right)
b_{m^{\prime }n^{\prime}}\left( {\bf k}^{\prime}\right), \label{of1}
\end{equation}
%======================================================================
\begin{equation}
{\cal F}^{\left(X\right)}_{m^{\prime }}\left( {\bf k}^{\prime}\right)
=\sum_{n^{\prime }}{\cal A}_{n^{\prime }}
\left( {\bf k}^{\prime }+{\bf q}\right) c_{m^{\prime }n^{\prime}}
\left( {\bf k}^{\prime}+{\bf q}\right), \label{of2}
\end{equation}
%======================================================================
which will specifically comprise the envelopes of the
functions of $\Gamma$ and $X_z$ states in ${\bf k}$ representation,
and we shall define the region $\Lambda_X$ such that the condition ${\bf k} \in \Lambda_X$
is satisfied for all ${\bf k+q} \in \tilde \Lambda_X$ (shift of the origin in
${\bf k}$ space). Now (\ref{expans2}) can have the form of an expansion of
$\Psi \left( {\bf r}\right)$ in terms of a complete set of Kohn-Luttinger functions
near the points $\Gamma$ and $X _z$:
%=======================================================================
\[
\Psi \left( {\bf r}\right) =\sum_{m^{\prime }}\int_{\Lambda_{\Gamma}}
{\cal F}^{\left(\Gamma\right)}_{m^{\prime }}\left( {\bf k}^{\prime }\right)
\chi^{\left(\Gamma\right)}_{m^{\prime }{\bf k}^{\prime}} \,
{\rm d}^3k^{\prime }+ \sum_{m^{\prime }}\int_{\Lambda_X}
{\cal F}^{\left(X\right)}_{m^{\prime }}\left( {\bf k}^{\prime } \right)
\chi^{\left(X\right)}_{m^{\prime }{\bf k}^{\prime}} \, {\rm d}^3k^{\prime }=
\]
%=======================================================================
\begin{equation}
=\sum_{v^{\prime }=\Gamma,X}\sum_{m^{\prime }}
\int_{\Lambda_{v^{\prime }}}{\cal F}^{\left(v^{\prime }\right)}_{m^{\prime }}
\left( {\bf k}^{\prime }\right) \chi^{\left(v^{\prime }\right)}_{m^{\prime }
{\bf k}^{\prime}} \, {\rm d}^3k^{\prime },  \label{expans-end}
\end{equation}
%=======================================================================
where the Kohn-Luttinger functions are
%=======================================================================
\[
\chi^{\left(\Gamma\right)}_{m{\bf k}}= e^{ i{\bf k r}}u_{m0} \equiv
e^{ i{\bf k r}}\phi^{\left(\Gamma\right)}_m , \quad
\chi^{\left(X\right)}_{m{\bf k}}=e^{ i{\bf k r}}e^{ i{\bf q r}}u_{m{\bf q}}
\equiv e^{ i{\bf k r}}\phi^{\left(X\right)}_m .
\]
%=======================================================================
Thus, the Fourier transforms of the envelope functions constructed above differ from the
usual ones \cite{lut-kohn} only in terms of the domains of their definition. In our case, these
are the regions $\Lambda_{\Gamma}$ (for states near the center of the Brillouin zone) and
$\Lambda_X$ (for states near $X _z$ point) rather than the complete first Brillouin zone.

For simplicity, below we shall also assume that $\phi^{(v)}_m$ are real. Taking the
following orthonormalization relationship for the Bloch functions:
%======================================================================
\begin{equation}
\int_{all \atop space}u^{*}_{n^{\prime}{\bf k}^{\prime}}
e^{-i{\bf k^{\prime}r}} u_{n{\bf k}}e^{i{\bf kr}} \, {\rm d}^3r=
\delta_{nn^{\prime}}\delta \left({\bf k-k^{\prime}}\right),
\label{bloch}
\end{equation}
%======================================================================
we obtain the required orthonormalization relationship
for the basis functions \cite{lut-kohn}:
%======================================================================
\begin{equation}
\int_{all \atop space}
\left(\chi^{\left(v^{\prime}\right)}_{n^{\prime}{\bf k}^{\prime}}\right)^{*}
\chi^{\left(v\right)}_{n{\bf k}} \, {\rm d}^3r=\delta_{vv^{\prime}}
\delta_{nn^{\prime}}\delta \left({\bf k-k^{\prime}}\right).
\label{basis}
\end{equation}
%-----------------------------------------------------------------------

\subsection{Multi-band System of ${\bf k \cdot p}$ Equations}

In the basis specified above the procedure for obtaining ${\bf k \cdot p}$ system
of equations is trivial, see \cite{lut-kohn}. Using the expansion (\ref{expans-end}) in
(\ref{schroedinger}), multiplying both sides of the equation by 
$(\chi^{\left(v\right)}_{m {\bf k}})^{*}$ and integrating over all ${\bf r}$
space, we obtain the following system of equations:
%======================================================================
\[
\left( \epsilon^{(v)} _m+\frac{{\hbar}^2{\bf k}^2}{2m_0}\right)
{\cal F}^{(v)}_m\left( {\bf k} \right) +\sum_{m^{\prime }}
\frac{{\hbar} {\bf p}^{(v)}_{mm^{\prime }}\cdot
{\bf k}}{m_0}{\cal F}^{(v)}_{m^{\prime }}\left( {\bf k}\right) + 
\]
\begin{equation}
+\sum_{v^{\prime }=\Gamma,X} \sum_{m^{\prime }}
\int_{\Lambda_{v^{\prime }}}{\cal M}^{(vv^{\prime })}_{mm^{\prime }}
\left( {\bf k},{\bf k}^{\prime }\right) {\cal F}^{(v^{\prime })}
_{m^{\prime }} \left( {\bf k}^{\prime }\right)\, {\rm d}^3 k^{\prime }
=\epsilon {\cal F}^{(v)}_m\left( {\bf k}\right) .  \label{kp}
\end{equation}
%======================================================================
Here $\epsilon^{(\Gamma)} _m =\epsilon_{m0}$ and $\epsilon^{(X)} _m
=\epsilon_{m{\bf q}}$,
%======================================================================
\[
{\bf p}^{(v)}_{mm^{\prime }}=
\left\langle m,v\mid {\bf p}\mid m^{\prime },v\right\rangle \equiv
\frac{\left( 2\pi \right) ^3}\Omega \int_{\rm cell}\phi^{\left(v\right)}_m
{\bf p}\phi^{\left(v\right)}_{m^{\prime}}\,{\rm d}^3r,
\]
%======================================================================
$\left\langle m,v\mid {\bf p}\mid m^{\prime },v^{\prime}\right\rangle =0$
for $v \neq v^{\prime}$, and $\Omega $ is the unit cell volume, and
%======================================================================
\[
{\cal M}^{(vv^{\prime })}_{mm^{\prime }}\left( {\bf k},{\bf k}^{\prime } \right) =
\int_{all \atop space} e^{-i\left( {\bf k}-{\bf k}^{\prime }\right){\bf r} }
\phi^{\left(v\right)}_m P\left( z\right)\delta U
\phi^{\left(v^{\prime}\right)}_{m^{\prime}} \, {\rm d}^3 r .
\]
%======================================================================

The matrix elements ${\cal M}^{(vv)}_{mm^{\prime }}
\left( {\bf k},{\bf k}^{\prime } \right)$ were analyzed in Ref.~\cite{we1}.
It was shown that the contribution of the perturbation potential can be divided into
``smooth'' and ``sharp'' components (the latter is exponentially small for
smooth perturbations on the scale $a$). The contribution of the sharp component is a
correction to that of the smooth component (for the case when the width of the
heterostructure layers is much greater than $a$), and it can be written in the
form of converging series in powers of $(k_z-k_z^{\prime })$. We will use the
effective-mass approximation with spatially independent effective-mass parameters
and we will only allow for the effects of abruptness of the heterojunctions in the first
order in terms of the parameter $a\bar k_z$ \cite{we1,we2}, where $\bar k_z$ is the
characteristic quasi-momentum of the state. We shall consider the intervalley elements
${\cal M}^{(vv^{\prime })}_{mm^{\prime }}
\left( {\bf k},{\bf k}^{\prime } \right)$, $v \ne v^{\prime}$ in greater detail and for
${\cal M}^{(vv)}_{mm^{\prime }} \left( {\bf k},{\bf k}^{\prime } \right)$, following
Ref.~\cite{we1,we2}, we obtain
%======================================================================
\[
{\cal M}^{(vv)}_{mm^{\prime }}\left( {\bf k},{\bf k}^{\prime }
\right) =\Biggl[ {\cal P}\left( k_z-k_z^{\prime }\right)
\delta U^{(v)}_{mm^{\prime }} +\frac 1 {2\pi} \sum_{j\ne0} \frac
{\left\langle m,v\mid \delta U e^{iK_jz}\mid m^{\prime },v\right\rangle}
{iK_j}\times
\]
\begin{equation}
\times \left( \int\limits_{-d}^d P^{\prime }(z) e^{ -iK_jz } {\rm d}z
-e^{-i(k_z-k_z^{\prime })L}
\int\limits_{-d}^dP^{\prime }(z) e^{ iK_jz } {\rm d}z \right) \Biggr]
\delta \left( {\bf k}_{\|}-{\bf k}_{\|}^{\prime }\right). \label{intra}
\end{equation}
%======================================================================
Here we introduce the notation: ${\cal P}(k_z)$ is the Fourier transform of the function
$P(z)$; the matrix element $\delta U^{(v)}_{mm^{\prime }}=
\left\langle m,v\mid \delta U \mid m^{\prime },v\right\rangle$;
$K_j = ( 4\pi /a )j$, and $j$ is an integer; $P^{\prime }(z)={\rm d}P(z)/{\rm d}z$;
and ${\bf k}_{\|}=(k_x,k_y,0 )$. We used the symmetry $P(z)=P(L-z)$, and also the fact that
$e^{ iK_jL }=1$. Here, we give all the matrix elements (\ref{intra}) required subsequently.
We assign the indices `{\rm w}', `{\rm u}' and `{\rm v}' to the conduction-band states
$\Gamma_1$, $X_1$ and $X_3$, respectively, and omit the valley indices since the band index
in our approximation now uniquely defines the state. Using symmetry concepts we obtain
%======================================================================
\[
{\cal M}_{ss}\left( {\bf k},{\bf k}^{\prime }
\right) =\Biggl[ {\cal P}\left( k_z-k_z^{\prime }\right)
\delta U_{ss} +\frac 1 {2\pi} d_{ss}\left( 1+
e^{-i(k_z-k_z^{\prime })L} \right) \Biggr]
\delta \left( {\bf k}_{\|}-{\bf k}_{\|}^{\prime }\right),
\]
%======================================================================
where $s=\rm w,\,u,\,v$; and the parameters $d_{ss}$ are determined as follows:
%======================================================================
\[
d_{ss} = -\sum_{j\ne0}
\frac{\left\langle s\mid \delta U \cos{(K_jz)}\mid s\right\rangle}{K_j}
\int\limits_{-d}^d P^{\prime }(z) \sin{(K_jz)} {\rm d}z.
\]
%======================================================================
The potential also makes a contribution to the direct interaction of $X_1$ and $X_3$ states:
%======================================================================
\[
{\cal M}_{\rm uv}\left( {\bf k},{\bf k}^{\prime } \right)
=\frac 1 {2\pi} d_{\rm uv}\left( 1-e^{-i(k_z-k_z^{\prime })L} \right)
\delta \left( {\bf k}_{\|}-{\bf k}_{\|}^{\prime }\right),
\]
%======================================================================
where
%======================================================================
\[
d_{\rm uv} = \sum_{j\ne0} \frac{\left\langle {\rm u}\mid
\delta U \sin{(K_jz)}\mid {\rm v}\right\rangle}{K_j}
\int\limits_{-d}^d P^{\prime }(z) \cos{(K_jz)}\, {\rm d}z.
\]
%======================================================================

We shall now consider the most interesting to us intervalley matrix elements
${\cal M}^{(vv^{\prime})}_{mm^{\prime }}$, $v \ne v^{\prime}$:
%======================================================================
\[
{\cal M}^{(\Gamma X)}_{mm^{\prime }}\left( {\bf k},{\bf k}^{\prime }
\right) =\int_{all \atop space} P\left( z\right)
e^{-i\left( {\bf k}-{\bf k}^{\prime }-{\bf q}\right){\bf r} }
u_{m0}\delta U u_{m^{\prime}{\bf q}}\, {\rm d}^3 r .
\]
%======================================================================
We shall analyze this matrix element using the same method which yielded (\ref{intra}).
We shall use an expansion of the periodic function $u_{m0}\delta U u_{m^{\prime}{\bf q}}$
as a Fourier series which gives
%======================================================================
\begin{equation}
{\cal M}^{(\Gamma X)}_{mm^{\prime }}\left( {\bf k},{\bf k}^{\prime }
\right) =\sum_l C^{m\left( \Gamma \right) m^{\prime }\left( X\right) }_l
{\cal P}\left( k_z-k_z^{\prime}-q_z+K_{zl}\right) \delta \left( {\bf k}_{\|}%
-{\bf k}^{\prime}_{\|}+{\bf K}_{\|l} \right), \label{inter1}
\end{equation}
%======================================================================
where ${\bf K}_l \equiv (K_{zl},{\bf K}_{\|l})$ are the vectors of the reciprocal
lattice, and
%======================================================================
\[
C^{m\left( \Gamma \right) m^{\prime }\left( X\right) }_l =
\frac{\left( 2\pi \right) ^3}\Omega \int_{\rm cell}
u_{m0}\delta U e^{i{\bf K}_l {\bf r}} u_{m^{\prime}\bf q} {\rm d}^3r \equiv
\left\langle m,\Gamma \mid \delta U e^{i\left({\bf K}_l-{\bf q}\right)
{\bf r}} \mid m^{\prime },X\right\rangle.
\]
%======================================================================
As was shown in Ref.~\cite{we2}, for the region where $\left| k_x\right|
+\left| k_y\right| <\pi /a$, whose size is fairly large for our purposes
[we obviously used this constraint in the derivation of (\ref{intra})], only the vectors of
the reciprocal lattice with ${\bf K}_{\|l}=0$ will contribute to (\ref{inter1}):
%======================================================================
\[
{\cal M}^{(\Gamma X)}_{mm^{\prime }}\left( {\bf k},{\bf k}^{\prime } \right)
=\sum_{j} \left\langle m,\Gamma \mid \delta U
e^{i\left( K_j-q_z\right) z}\mid m^{\prime },X \right\rangle
{\cal P}\left( k_z-k_z^{\prime}+K_j-q_z \right)
\delta \left( {\bf k}_{\|}-{\bf k}^{\prime}_{\|}\right) =
\]
\begin{equation}
=\sum_{j=\pm 1,\pm 3, \pm 5, \dots} \left\langle m,\Gamma \mid \delta U
e^{i\frac {2\pi}ajz}\mid m^{\prime },X \right\rangle
{\cal P}\left( k_z-k_z^{\prime}+\frac {2\pi}aj \right)
\delta \left( {\bf k}_{\|}-{\bf k}^{\prime}_{\|}\right) . \label{inter2}
\end{equation}
%======================================================================
We shall analyze the function ${\cal P}(k_z-k_z^{\prime}+(2\pi /a)j )$,
contained in (\ref{inter2}) in greater detail:
%======================================================================
\[
{\cal P}\left( k_z-k_z^{\prime}+\frac {2\pi}aj \right)=
\frac 1{2\pi i}\frac 1{k_z-k_z^{\prime }+\frac {2\pi}aj}
\int\limits_{-\infty }^{+\infty }P^{\prime }\left( z\right)
e^{ -i\left( k_z-k_z^{\prime }+\frac {2\pi}aj \right) z}\,{\rm d}z=
\]
\[
=\frac 1{2\pi i}\frac 1{k_z-k_z^{\prime }+\frac {2\pi}aj}
\Biggl( \int\limits_{-d }^d P^{\prime }\left( z\right)
e^{ -i\frac {2\pi}aj z} e^{ -i\left( k_z-k_z^{\prime }\right) z}\,{\rm d}z-
\]
\[
-e^{ -i\left( k_z-k_z^{\prime }\right) L}e^{ -i\frac {2\pi}aj L}
\int\limits_{-d }^d P^{\prime }\left( z\right) e^{i\frac {2\pi}aj z}
e^{i\left( k_z-k_z^{\prime }\right) z}\,{\rm d}z \Biggr),
\]
%======================================================================
where we again used the property $P(z)=P(L-z)$. We now expand
$(k_z-k_z^{\prime }+(2\pi/a)j)^{-1}$,
$\exp( -i( k_z-k_z^{\prime }) z)$, and $\exp( i( k_z-k_z^{\prime }) z)$ as series in
powers of $( k_z-k_z^{\prime })$, where the convergence of the first series is ensured by the
property $\mid k_z-k_z^{\prime }\mid <2\pi/a$, since ${\bf k} \in \Lambda_{\Gamma}$ and
${\bf k}^{\prime} \in \Lambda_X$. We only retain the first terms of these expansions.
This gives a good approximation for $\bar{k}_z \ll 2\pi /a$ and $\bar{k}_z \ll 1/(2d)$,
where $\bar{k}_z$ is the characteristic quasi-momentum of the state or the reciprocal
characteristic length of variation  of the envelope functions (\ref{of1}) and (\ref{of2}) in
${\bf r}$ representation (i.e., the envelope functions should vary weakly on scales of order $a$
and on scales of the order of the widths of the interface regions). Now, bearing in 
mind that for all values of the summation index $j$ in (\ref{inter2})
$\exp( -2\pi i j L/a)=(-1)^{\cal N}$, we obtain:
%======================================================================
\[
{\cal P}\left( k_z-k_z^{\prime}+\frac {2\pi}aj \right)=
\frac 1{2\pi i}\frac a{2\pi j}
\Biggl( \int\limits_{-d }^d P^{\prime }\left( z\right)
e^{ -i\frac {2\pi}aj z} \,{\rm d}z-
\]
\[
-(-1)^{\cal N} e^{ -i\left( k_z-k_z^{\prime }\right) L}
\int\limits_{-d }^d P^{\prime }\left( z\right) e^{i\frac {2\pi}aj z}
\,{\rm d}z \Biggr).
\]
%======================================================================
Here, there is a dependence of the effective potential on the number of monoatomic layers
$\cal N$, which mixes the valley states assigned to different points in ${\bf k}$ space. We
now write the matrix elements we require (we again drop the valley index):
%======================================================================
\[
{\cal M}_{\rm wu}\left( {\bf k},{\bf k}^{\prime } \right) =\frac 1 {2\pi}
d_{\rm wu}\left( 1+\left( -1\right)^{\cal N} e^{-i(k_z-k_z^{\prime })L}
\right) \delta \left( {\bf k}_{\|}-{\bf k}_{\|}^{\prime }\right),
\]
%=======================================================================
where
%=======================================================================
\begin{equation}
d_{\rm wu} = -\sum_{j=\pm 1,\pm 3, \pm 5, \dots} \frac a{2\pi j}
\left\langle {\rm w}\mid \delta U \cos{(\frac {2\pi}ajz)}
\mid {\rm u}\right\rangle
\int\limits_{-d}^d P^{\prime }(z) \sin{(\frac {2\pi}ajz)}\,{\rm d}z;
\label{dwu}
\end{equation}
%======================================================================
and
%======================================================================
\[
{\cal M}_{\rm wv}\left( {\bf k},{\bf k}^{\prime } \right) =\frac 1 {2\pi}
d_{\rm wv}\left( 1-\left( -1\right)^{\cal N} e^{-i(k_z-k_z^{\prime })L}
\right) \delta \left( {\bf k}_{\|}-{\bf k}_{\|}^{\prime }\right),
\]
%=======================================================================
where
%=======================================================================
\begin{equation}
d_{\rm wv} = \sum_{j=\pm 1,\pm 3, \pm 5, \dots} \frac a{2\pi j}
\left\langle {\rm w}\mid \delta U \sin{(\frac {2\pi}ajz)}
\mid {\rm v}\right\rangle
\int\limits_{-d}^d P^{\prime }(z) \cos{(\frac {2\pi}ajz)}\,{\rm d}z.
\label{dwv}
\end{equation}
%======================================================================
It can be seen that the abrupt potential of the heterointerfaces not only ensures
$\Gamma_1$-$X_3$ \cite{ando,ivchenko,foreman}, but also $\Gamma_1$-$X_1$ interaction
\cite{klipstein}.

We have obtained a multi-band ${\bf k \cdot p}$ system of equations and explicit
expressions for all the elements of this system required subsequently. We shall now obtain
the system of three equations for the strongly interacting states $\Gamma_1$, $X_1$, and $X_3$.
%-----------------------------------------------------------------------

\subsection{Elimination of Remote Bands and Transition to ${\bf r}$ Space}

In the effective-mass approximation with spatially independent effective-mass parameters,
the approximate unitary transformation procedure which eliminates the influence of remote
bands in the required order of perturbation theory is performed by a standard
method \cite{lut-kohn}, and is not given here. We obtain an integral
system of equations in ${\bf k}$ representation. The aim of the present study is to
simplify the final results as far as possible which may be achieved if the corresponding
equations are differential. The problem of the accuracy of the envelope function method
which occurs on transition from integral to differential equations and also when the
remote bans are eliminated has already been discussed in Refs.~\cite{we1,we2}.
We shall summarize the constraints imposed on the accuracy of the envelope
function method for this case. First, we need to determine the effective radius of
the region in ${\bf k}$ space for which the system of equations for the envelope
functions is correct (we shall call this ${\bf k}$ region the fundamental region).
Whereas the multi-band system of equations (\ref{kp}) is valid for all ${\bf k} \in
\Lambda_v$ and ${\bf k}^{\prime} \in \Lambda_{v^{\prime}}$, the unitary transformation
of this system which eliminates the remote bands can, in principle, reduce the dimensions
of the fundamental ${\bf k}$ regions. This is easily understood from the following. The
spectrum of states of a bulk semiconductor $\epsilon^{({\bf k}_0)}_{n}({\bf k})$ near the
point ${\bf k}_0$ in a band numbered $n$ may be represented as a series in powers of
${\bf k}$ (for degenerate states the spectrum is determined by diagonalizing the matrix whose
elements are such series). The series has a finite radius of convergence $R_0$, which is
determined by the strength of the ${\bf k\cdot p}$ interaction with the remote bands. This
radius can be estimated as $R_0=m_0 \bar E_g/(2\hbar \bar P)$, where $\bar E_g$ and
$\bar P$ are the characteristic values of the interband energy at point ${\bf k}_0$
and the interband matrix element of the momentum, respectively. States having quasi-momenta
which do not belong to the fundamental region cannot be correctly taken into account in
the transformed equation and should be neglected. We denote the corresponding radii
of the fundamental ${\bf k}$ regions as $R^{(\Gamma)}_0$ and $R^{(X)}_0$ for
conduction-band states near the points $\Gamma$ and $X$, respectively.

Secondly, on changing from ${\bf k}$ to ${\bf r}$ representation, integration
is performed over regions of ${\bf k}$ space of finite dimensions which prevents us from
direct obtaining differential equations in ${\bf r}$ space. The local approximation
formula involves replacing the finite regions of ${\bf k}$ space by infinite ones.
Since the heterostructure potential is not smooth and the envelope functions obtained
(or their derivatives) can vary appreciably on scales of the order of $a$, this procedure
does not give an exponentially small error as in the case of smooth perturbations,
but an error which is only small in terms of power (for this analysis it is convenient to
consider the limiting case of a mathematically abrupt potential and then the envelope
functions or their derivatives obtained as a result of the local approximation
may have a discontinuity). Since the perturbation-theory series used for the unitary
transformation is a power series, it is important to avoid the inclusion of
extra-accuracy terms. For the simple single-valley case ($\Gamma$ states) analyzed in detail
in Ref.~\cite{we2}, the error of the method is of the order of
$(\bar k^{(\Gamma)}_z/R^{(\Gamma)}_0 )^M$, where $1/\bar k^{(\Gamma)}_z$ is
the characteristic length of variation of the corresponding envelope function and the
exponent $M$ is a measure of the smoothness of the latter (for an isolated heterojunction
or a fairly wide quantum well $M = 3$). In this case, we need to determine two characteristic
quasi-momenta $\bar k^{(\Gamma)}_z$ and $\bar k^{(X)}_z$ for $\Gamma$ and $X$ states.
In the zero-order approximation, which is acceptable for estimating the accuracy of the
method, the $\Gamma$ and $X$ states do not interact and the constraint associated with the
transition to differential equations is determined by the error
$(\bar k^{(\Gamma)}_z/R^{(\Gamma)}_0 )^3$ for $\Gamma$ states (we shall assume that the
layer width $L$ is sufficiently large so that $\bar k^{(\Gamma)}_zL$~{\amsfnt\char"26}~$1$).
For $X$ states, the situation is slightly more complex: the system of equations for these
contains both second derivatives of the envelope functions with respect to $z$ and first
derivatives as a result of ${\bf k \cdot p}$ interaction between $X_1$ and $X_3$ bands,
where the effect of this interaction may be comparable with the contribution of terms which are
quadratic with respect to the momentum operator. This means that in the ``worst'' case,
the accuracy of the local approximation for $X$ states is limited by the error
$(\bar k^{(X)}_z/R^{(\Gamma)}_0 )^2$. All these factors allow us to consider the
corrections which appear as a result of the abruptness of the change in the heterointerface
potential which are small as small are the parameters $a\bar k^{(\Gamma)}_z$ and
$a\bar k^{(X)}_z$. Here we also note that splitting the region $\Lambda_0$ into the subregions
$\Lambda_{\Gamma}$ and $\tilde \Lambda_X$ should be performed so that $R^{(\Gamma)}_0$ and
$R^{(X)}_0$ are not larger than the radii of the regions $\Lambda_{\Gamma}$ and $\Lambda_X$,
respectively, and then $R^{(\Gamma)}_0$ and $R^{(X)}_0$, not the radii $\Lambda_{\Gamma}$ and
$\Lambda_X$ appear in the expressions to estimate the accuracy of the local approximation
(this was implied above).

As a result, the required system of differential equations for the transformed envelope
functions $\tilde F_m \left( {\bf r} \right)$, where $m = {\rm w,\,u,\,v}$, has the following
form for the strongly interacting $\Gamma_1$, $X_1$, and $X_3$ states:
%==========================================================================
\begin{equation}
\sum_{m^{\prime}=\rm w,\,u,\,v} \left( T_{mm^{\prime}} + V_{mm^{\prime}}
\left( z\right) \right) \tilde F_{m^{\prime}}\left( {\bf r} \right)
= \epsilon \tilde F_m \left( {\bf r} \right). \label{system}
\end{equation}
%==========================================================================
Here, ${\bf T}$ and ${\bf V}$ are ($3\times3$) matrices of the kinetic and
potential energies. The form of the matrix of the effective kinetic energy operator
is known \cite{ivch-phot}:
%==========================================================================
\[
\bf T =\left( \matrix{
\frac{{\bf p}^2}{2m_{\rm w}}&0&0 \cr
0&\frac{{\bf p}_{\|}^2}{2m_{\rm u\|}} + \frac {{\rm p}_z^2}{2m_{\rm u\bot}}&
\frac{ ({\rm p}_z)_{\rm uv} {\rm p}_z}{m_0} +\gamma{\rm p}_x{\rm p}_y\cr
0&\frac{ ({\rm p}_z)_{\rm vu} {\rm p}_z}{m_0} +\gamma{\rm p}_x{\rm p}_y&
\frac {{\bf p}_{\|}^2} {2m_{\rm v\|}} + \frac {{\rm p}_z^2}{2m_{\rm v\bot}}
} \right),
\]
%==========================================================================
where $m_{\rm w}$ is the effective mass for $\Gamma$ conduction band states,
$m_{l\|}$ and $m_{l\bot}$ are the longitudinal and transverse effective masses for
$l$ band ($l = \rm u,\,v$) at point $X$; and the bulk parameter $\gamma$, in particular,
determines the magnitude of the linear photogalvanic effect \cite{ivch-phot}.
Before giving the form of the matrix of the potential energy operator, we go over from the
function $P(z)$ to the function $\Theta (z) -\Theta (z-L)$ merely for the reasons of
convenience \cite{we2}:
%======================================================================
\begin{equation}
P \left( z \right) \approx \Theta \left( z\right) -\Theta \left( z-L\right)
+ \rho_0\left( \delta \left( z\right) +\delta \left( z-L\right) \right),
\label{theta}
\end{equation}
%======================================================================
where
%======================================================================
\[
\rho _0=\int\limits_{-d}^dP \left( z\right) {\rm d}z -d . 
\]
%======================================================================
Now the matrix of the effective potential energy operator ${\bf V}$
may be expressed as the sum of three matrices: ${\bf V}_1 + {\bf V}_2 + {\bf V}_3$.
The diagonal matrix ${\bf V}_1$ corresponds to the standard (bulk) effective-mass
approximation \cite{lut-kohn}:
%==========================================================================
\[
\left( V_1\right)_{ss^\prime}=\{ \epsilon _{ss^\prime} + \delta U_{ss^\prime}
\left[ \Theta \left( z\right) -\Theta \left( z-L\right)\right] \} \delta_{ss^\prime}.
\]
%==========================================================================
As a result of the abruptness of the heterointerface potential, the matrix ${\bf V}_2$
contains both intravalley contributions and contributions which mix $X_1$ and $X_3$
states (all the parameters $d_{mm^{\prime}}$ are real):
%==========================================================================
\[
{\bf V}_2=\left( \matrix{
\tilde d_{\rm ww}\left( \delta (z)+\delta (z-L) \right)&0&0\cr
0&\tilde d_{\rm uu}\left( \delta (z)+\delta (z-L) \right)&
d_{\rm uv}\left( \delta (z)-\delta (z-L) \right)\cr
0&d_{\rm uv}\left( \delta (z)-\delta (z-L) \right)&
\tilde d_{\rm vv}\left( \delta (z)+\delta (z-L) \right)
} \right).
\]
%==========================================================================
Here, the parameters $\tilde d_{\rm ss}$ are related to $d_{\rm ss}$ as follows:
%=========================================================================
\[
\tilde d_{\rm ss} = d_{\rm ss} + \delta U_{\rm ss} \rho _0.
\]
%========================================================================
Finally, the matrix ${\bf V}_3$ contains contributions which mix $\Gamma$ and
$X_z$ states (also as a result of the abruptness of the heterointerface potential).
We give the non-zero elements of ${\bf V}_3$:
%==========================================================================
\begin{equation}
\left( V_3\right)_{\rm wu}=\left( V_3\right)_{\rm uw}=
d_{\rm wu}\left( \delta (z)+\left( -1\right)^{\cal N}\delta (z-L)\right),
\label{delta1}
\end{equation}
%==========================================================================
\begin{equation}
\left( V_3\right)_{\rm wv}= \left( V_3\right)_{\rm vw}=
d_{\rm wv}\left( \delta (z)-\left( -1\right)^{\cal N}\delta (z-L)\right).
\label{delta2}
\end{equation}
%==========================================================================

In our approximation, the expression linking the envelope functions and the complete wave
function has the usual form (we again drop the valley index):
%======================================================================
\[
\Psi \left( {\bf r}\right) =\sum_{m={\rm w,\,u,\,v}}
\left[ \tilde F_m \left( {\bf r} \right) \phi_m +
{\sum_n}^{\prime } \frac{ \hbar {\bf p}_{nm} \cdot
\left( \nbla \tilde F_m\left( {\bf r}\right) \right)}
{im_0 \left( \epsilon _m -\epsilon _n\right)} \phi_n \right].
\]
%-----------------------------------------------------------------------

For an arbitrary number of heterojunctions at $z=z_j$ numbered by the index $j$,
the nonzero elements of ${\bf V}_3$ may be written in the following form:
%==========================================================================
\[
\left( V_3\right)_{\rm wu}=\left( V_3\right)_{\rm uw}=
\sum_j e^{iqz_j}d^{(j)}_{\rm wu}\delta (z-z_j),
\]
%==========================================================================
\[
\left( V_3\right)_{\rm wv}= \left( V_3\right)_{\rm vw}=
\sum_j e^{iqz_j}d^{(j)}_{\rm wv}\delta (z-z_j),
\]
%==========================================================================
where $q$ is the distance in ${\bf k}$ space between the centers of the valleys under
consideration (in our case, $q = (0,0,2\pi /a)$). Since the parameters $d^{(j)}_{\rm wu}$
and $d^{(j)}_{\rm wv}$ depend not only on the heterojunction materials but also on
the microscopic structure of the interface, these must generally be determined for each
interface separately.

\section{Discussion of Results and Conclusions}

Using ${\bf k\cdot p}$ formalism, we have constructed a generalization of the
method of envelope functions suitable to describe the interaction of $\Gamma$ and $X_z$
states in (001) III-V nanostructures formed from related lattice-matched semiconductors.
In the derived system of equations (\ref{system}), mixing of the states of different valleys
is determined by the heterosurface effective potentials (\ref{delta1}) and (\ref{delta2}),
similar to those introduced phenomenologically in Ref.~\cite{liu}. The system
(\ref{system}) contains information on the number $\cal N$ of monoatomic layers of the structure.
An oscillatory dependence of the effective intervalley mixing strength on
$\cal N$ was obtained, in particular, in Ref.~\cite{lu}, and was also introduced from
symmetry concepts (in terms of the boundary conditions for the envelope functions) in
Ref.~\cite{ivchenko}. However, in addition to this, there are some difference between the
results of the present study and the results of other authors. The most important of these is
the appearance of direct interaction between $\Gamma_1$ and $X_1$ states whose strength is
determined by the parameter $d_{\rm wu}$. This was predicted recently in
Refs.~\cite{klipstein,wegx,disser}. The strength of $\Gamma_1$-$X_1$ interaction, i.e.,
the value of the constant $d_{\rm wu}$, depends strongly on the structure of the
heterointerface on atomic scales. In simplified models, such as the simplest variants of the
tight-binding method, this interaction may be absent. It is clear from (\ref{dwu}) that in the
hypothetical case of mathematically abrupt heterojunctions when $P(z)=\Theta (z)-\Theta (z-L)$,
we in fact find $d_{\rm wu}=0$. The conclusion reached in Ref.~\cite{foreman} that
$d_{\rm wu} \ll d_{\rm wv}$ was specifically a consequence of the selected heterojunction
model for which the simplest case is a mathematically abrupt jump (see also
Ref.~\cite{klipstein}). In general, there is no basis for assuming that
$d_{\rm wu}$ differs substantially from $d_{\rm wv}$, and since $X_1$ band is lower in
energy than $X_3$ one, the existence of $\Gamma_1$-$X_1$ interaction may be very significant
for interpreting experiments.

In Ref.~\cite{we2} we considered a more complex model of the potential of an ideal
heterostructure and, specifically, took into account a periodic coordinate dependence of
the heterostructure form-factor in the $(001)$ plane. We showed that this complication
merely leads to renormalization of some parameters obtained using (\ref{lattice}).
However, whereas it was found using the simple form-factor model that the mixing strength of
the heavy and light holes at the center of the 2D Brillouin zone was higher for abrupt
heterojunctions than for structures with a continuously varying composition, this conclusion was
not obtained using the more complex form-factor model. We could merely conclude that the strength
of the light hole-heavy hole mixing depended on the structure of the transition region of the
heterointerfaces. Having made a similar analysis for our case, we can easily show that
$\Gamma$-$X$ mixing strength in fact depends strongly on how abruptly (on scales of the order
$a$) the chemical composition at the heterojunction varies. This conclusion which follows
directly from expressions (\ref{dwu}) and (\ref{dwv}) for the parameters $d_{\rm wu}$ and
$d_{\rm wv}$ is quite natural since transitions with such a large (of the order of the
dimensions of the Brillouin zone) change in the quasi-momentum $k_z$ can only be achieved
by electron scattering at an atomically abrupt heterointerface (in formal terms the
Fourier components of the heterointerface potential with the wave vectors close to  $k_z=2\pi /a$
are responsible for $\Gamma$-$X$ transitions).

\section*{Acknowledgements}

This work was supported financially by the Russian Foundation for Basic Research
(project no. 99-02-17592), INTAS (grant no. 97-11475), and under the programs
"Physics of Solid-State Nanostructures" (project no. 99-1124) and "Surface Atomic Structures"
(project no. 3.1.99).
%%%%%%%%%%%%%%%%%%%%%%%%%%%%%%%%%%%%%%%%%%%%%%%%%%%%%%%%%%

\end{document}